\documentclass[epj]{svjour}
\usepackage{graphics}
\begin{document}
\def\dsdt{$\frac{d\sigma}{dt}$}
\def\beqn{\begin{eqnarray}}
\def\eeqn{\end{eqnarray}}
\def\barr{\begin{array}}
\def\earr{\end{array}}
\def\btab{\begin{tabular}}
\def\etab{\end{tabular}}
\def\bite{\begin{itemize}}
\def\eite{\end{itemize}}
\def\bcen{\begin{center}}
\def\ecen{\end{center}}

\def\eq{\begin{equation}}
\def\ee{\end{equation}}
\def\eqa{\begin{eqnarray}}
\def\eea{\end{eqnarray}}
\def\nn{\nonumber}
\def\psmu{P^{\prime \mu}}
\def\psnu{P^{\prime \nu}}
\def\ksmu{K^{\prime \mu}}
\def\pss{P^{\prime \hspace{0.05cm}2}}
\def\psf{P^{\prime \hspace{0.05cm}4}}
\def\kdagger{K\hspace{-0.3cm}/}
\def\ndagger{N\hspace{-0.3cm}/}
\def\psdagger{p'\hspace{-0.28cm}/}
\def\epssdagger{\varepsilon'\hspace{-0.28cm}/}
\def\epsdagger{\varepsilon\hspace{-0.18cm}/}
\def\pdagger{p\hspace{-0.18cm}/}
\def\xidagger{\xi\hspace{-0.18cm}/}
\def\qsdagger{q'\hspace{-0.3cm}/}
\def\qdagger{q\hspace{-0.2cm}/}
\def\keldagger{k\hspace{-0.2cm}/}
\def\ksdagger{k'\hspace{-0.3cm}/}
\def\q2dagger{q_2\hspace{-0.35cm}/\;}
\def\qqs{q\!\cdot\!q'}
\def\lls{l\!\cdot\!l'}
\def\lp{l\!\cdot\!p}
\def\lps{l\!\cdot\!p'}
\def\lsp{l'\!\cdot\!p}
\def\lsps{l'\!\cdot\!p'}
\def\lqs{l\!\cdot\!q'}
\def\pps{p\!\cdot\!p'}
\def\psqs{p'\!\cdot\!q'}
\def\epsp{\varepsilon'\!\cdot\!p}
\def\epsps{\varepsilon'\!\cdot\!p'}
\def\epsl{\varepsilon'\!\cdot\!l}
\def\epsls{\varepsilon'\!\cdot\!l'}

\title{Generalized sum rules of the nucleon in the constituent quark model}
\author{M. Gorchtein\inst{1}  \and D. Drechsel\inst{2} \and M. Giannini\inst{1} \and E. Santopinto\inst{1} 
}                     
\institute{Universit\,a di Genova, Sezione INFN di Genova, via Dodecaneso 33, 16142 Genova (Italy) \and Institut f\"ur Kernphysik, Universit\"at Mainz, Staudingerweg 45, 55099 Mainz (Germany)}
\date{Received: 05.08.2003 / Revised version: 05.08.2003}
%
\abstract{
The sum rules serve a 
powerful tool to study the nucleon structure by providing a bridge 
between the statical properties of the nucleon (such as electrical charge, 
and magnetic moment) and the dynamical properties (e.g. the transition 
amplitudes to excited states) in a wide range of energy and momentum 
transfer $Q^2$. We study the generalized sum rules of the nucleon in the 
framework of the constituent quark model. We use two different CQM, the one
with the hypercentral potential \cite{hccqm1}, \cite{hccqm2}, \cite{hccqm3}, 
and with the harmonic oscillator potential \cite{KI},
both with only few parameters fixed to the baryonic spectrum. 
We confront our results to the model independent sum rules and to the 
predictions of the phenomenological MAID \cite{MAID} model and find that 
in all the cases 
considered, in the intermediate $Q^2$ range (0.2-1.5 GeV$^2$), both CQM 
models provide a good description of the sum rules on the neutron.
} 
\maketitle

\section{Introduction}
\label{intro}

In the recent years, precise measurements of single and double polarization
observables for the photo- and electorabsorption have become possible. 
The inclusive cross section for the process 
$(ep\rightarrow eX)$ can be written in terms of the four partial cross 
sections,

\beqn
\frac{d\sigma}{d\Omega dE'}&=&\Gamma_V
\left[
\sigma_T\,+\,\sigma_L\,-\,hP_x\sqrt{2\epsilon(1-\epsilon)}\sigma_{LT}
\right.\nn\\
&&\;-\,\left.
hP_z\sqrt{1-\epsilon^2}\sigma_{TT}
\right]\;,
\label{eq:crosssec}
\eeqn
\noindent
with $\sigma_T=\frac{\sigma_{1/2}+\sigma_{3/2}}{2}$,  
$\sigma_{TT}=\frac{\sigma_{1/2}-\sigma_{3/2}}{2}$,  
the virtual photon flux factor $\Gamma_V\;=\;\frac{\alpha_{em}}{2\pi^2}\frac{E'}{E}\frac{K}{Q^2}\frac{1}{1-\epsilon}$, and the photon polarization
$\epsilon\;=\;\frac{1}{1+2(1+\nu^2/Q^2)\tan^2(\Theta/2)}$, where $E(E')$ 
denote the initial (final) electron energy, $\nu=E-E'$ the
energy transfer to the target, $\Theta$ the electron c.m. scattering angle, 
and $Q^2=4EE'\sin^2(\Theta/2)$ the four momentum transfer. The virtual photon 
spectrum normalization factor is chosen to be $K\,=\,\frac{s-M^2}{2M}$, 
$h=\pm 1$ refers to the electron helicity,
while $P_z$ and $P_x$ are the components of the target polarization.
For a general and complete consideration of the nucleon sum rules we readress
the reader to the review \cite{reviewdr}.

\section{Constituent quark model}
We study the generalized sum rules for the nucleon within a 
hyper central constituent quark model (HCCQM) previously reported in 
\cite{hccqm1}, \cite{hccqm2}, \cite{hccqm3}. The model based on the lattice
QCD inspired potential of the form 
$V(x)\,=\,-\frac{\alpha}{x}+\beta x+V_{hyp}$ and allows
for a consistent description of the baryonic spectrum with a minimal number of 
parameters. Furthermore, to display the 
model dependence of a CQM calculation, we list also the results within the
CQM with a harmonic oscillator (HO) potential. Within this 
model, we for the first time report a calculation of the longitudinal 
amplitudes and their contribution to the sum rules, details of which will be 
reported in an upcoming article.

The electromagnetic transition helicity amplitudes are defined as
\beqn
A_{1/2}&=&-\frac{e}{\sqrt{2\omega}}
<R,{1\over 2}|J_+|N,-{1\over 2}>\;,
\nn\\
A_{3/2}&=&-\frac{e}{\sqrt{2\omega}}
<R,{3\over 2}|J_+|N,{1\over 2}>\;,
\nn\\
S_{1/2}&=&\frac{e}{\sqrt{2\omega}}
<R,{1\over 2}|\rho|N,{1\over 2}>\;,
\eeqn
where ${1\over 2}$, ${3\over 2}$ stands for the spin projection of the 
initial (nucleon) and final (resonance) hadronic state, and the definition was
used, 
$J_+\,\equiv\,\vec{\varepsilon}^+\cdot \vec{J}\,=\,-\frac{J_x+iJ_y}{\sqrt{2}}$.

We will present the results for the sum rules within the $zero-width$ 
approximation where one has for the contribution of a single resonance $R$ to 
the partial cross sections
\beqn
\sigma_L^R&=&2\pi\delta(\nu-\nu_R)\frac{Q^2}{q^2_R}|S_{R,1/2}|^2\nn\\
\sigma_{LT}^R&=&\sqrt{2}\pi\delta(\nu-\nu_R)\frac{Q}{q_R}
\left(S_{R,1/2}\cdot A_{R,1/2}\right)\nn\\
\sigma_{1/2,3/2}^R&=&2\pi\delta(\nu-\nu_R)|A_{R,1/2,3/2}|^2\;,
\eeqn
with $\nu_R\,=\,\frac{M_R^2-M^2+Q^2}{2M}$, $q_R\,=\,\sqrt{\nu_R^2+Q^2}$, 
and $M_R$ the resonance mass.

\section{Sumrules for the forward polarizabilities of the proton}
We start with the Baldin sum rule which relates the sum of the electromagnetic
polarizabilities to the integral over the total photoabsorption cross section,
\beqn
\alpha(Q^2)\,+\,\beta(Q^2)\;=\;\frac{1}{2\pi^2}
\int_{\nu_0}^\infty
\frac{K}{\nu}\frac{\sigma_{T}(\nu, Q^2)}{\nu^2}d\nu\;,
\eeqn
where $\nu_0=m_\pi+\frac{m_\pi^2+Q^2}{2M}$ is the pion production threshold. 

\begin{figure}
\resizebox{0.4\textwidth}{!}{%
  \includegraphics{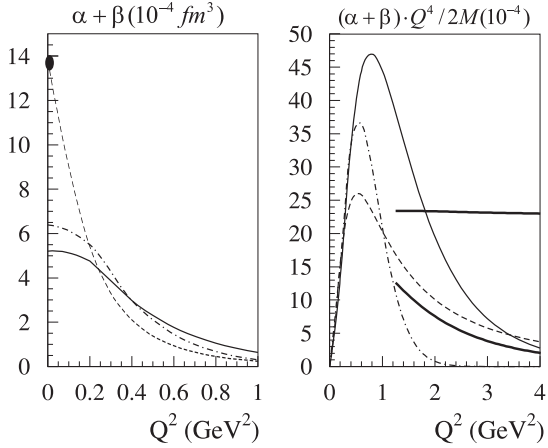}
}
\caption{Results for the sum of the proton polarizabilities $\alpha + \beta$
calculated in HCCQM (solid line), HO model (dashed-dotted line)
in comparison with MAID (dashed line). The solid circle corresponds to the 
Baldin sum rule value at $Q^2=0$ \cite{MAMI}.}
\label{fig:baldin}
\end{figure}

As it can be seen from Fig. \ref{fig:baldin}, both constituent quark models 
fall short at $Q^2=0$ by a factor of 3, which is a consequence of lacking
the large contribution of pion production.
However, starting from $Q^2=0.2$ GeV$^2$ all three models give similar results.

\begin{figure}
\resizebox{0.4\textwidth}{!}{%
  \includegraphics{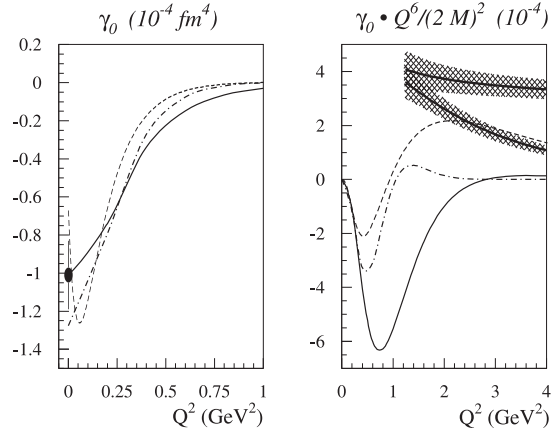}
}
\caption{Results for $\gamma_0$ on proton. Notation as in Fig. 
\ref{fig:baldin}. The data point at $Q^2=0$ is from \cite{MAMI}.}
\label{fig:gamma0}
\end{figure}

We next turn to the sum rules with the helicity flip cross section 
$\sigma_{TT}$. In Fig. \ref{fig:gamma0}, we show the results for the forward 
spin polarizability,
\beqn
\gamma_0(Q^2)\;=\;\frac{1}{2\pi^2}\int_{\nu_0}^\infty
\frac{K}{\nu}\frac{\sigma_{TT}(\nu, Q^2)}{\nu^3}d\nu
\eeqn

In the case of $\gamma_0$, the small value phenomenologically comes about 
due to a strong 
cancellation of a large negative contribution of the $\Delta (1232)$ 
resonance, and a large positive contribution of near threshold pion production.
Though the latter is not present in neither of the two presented quark model 
calculations, both do surprisingly well for this sum rule, as can be seen in 
Fig. \ref{fig:gamma0}, since almost all the transition helicity amplitudes in
a CQM lack strenght, as compared to the phenomenological analysis. Therefore,
the fact that the quark model results are consistent with the results of MAID
in the shown range of $Q^2$ should be seen rather as a coincidence. It is 
interesting to note that, due to the characteristical for the HO potential 
gaussian form factors, the HO model closely reproduces the slope of the MAID 
curve.

\section{Generalized GDH sum rule}
The GDH sum rule relates the anomalous magnetic moment of the nucleon 
to the integral over its excitation spectrum,
\beqn
-\frac{\kappa^2}{4}\;=\;\frac{M^2}{2\pi e^2}
\int_{\nu_0}^\infty
d\nu\frac{\sigma_{1/2}-\sigma_{3/2}}{\nu}\;,
\eeqn
thus providing a test of a quark model, since both left and right hand sides 
of this sum rule can be calculated. One of the possible generalizations of 
this integral to the case of finite $Q^2$ is
\beqn
I_A(Q^2)&=&\frac{M^2}{\pi e^2}
\int_{\nu_0}^\infty
d\nu\frac{K}{\nu}\frac{\sigma_{TT}(\nu,Q^2)}{\nu}\;,
\eeqn

\begin{figure}
\resizebox{0.45\textwidth}{!}{%
  \includegraphics{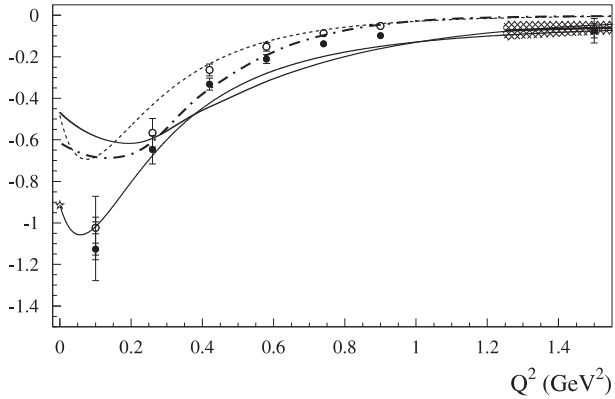}
}
\caption{Results for the generalized GDH integral $I_A^n$. Notation as in 
Fig. \ref{fig:baldin}. 
The thin solid line corresponds to the experimental fit, 
as described in text. The solid line starting from 1.25 GeV$^2$ corresponds 
to the evaluation of the GDH integral using the data on the DIS structure 
functions (for the details, see \cite{reviewdr}). The solid star represents 
the sum rule value at $Q^2=0$. The data points are from
\cite{HERMES} (solid squares) and
\cite{IAnJLAB} (solid (full result) and open (resonant part with 
$W\leq 2$ GeV) circles).}
\label{fig:ian}
\end{figure}

In Fig. \ref{fig:ian}, we show the results for the GDH integral $I_A$ on 
the neutron. As one can see, the sum rule at the real photon point
is not obeyed in either quark model. 

Starting from $Q^2=0.3$ GeV$^2$, the 
HCCQM practically follows the experimental fit, which assumes the following 
form (for details, see \cite{reviewdr}):
\beqn
I_A(Q^2)&=&I_A^{res}(Q^2)\,+\,2M^2\Gamma_A^{as}
\left[
\frac{1}{Q^2+\mu^2}\,-\,\frac{c\mu^2}{(Q^2+\mu^2)^2}
\right]\;,
\nn\\
c&=&1\,+\,\frac{\mu^2}{2M^2\Gamma_A^{as}}
\left[
\frac{\kappa^2}{4}\,+\,I_A^{res}(0)
\right]\;,
\eeqn
with $\mu=m_\rho$ and the resonance part as calculated with MAID.
Due to the characteristical 
HO gaussian form factors having a more steep $Q^2$ dependence, the HO model
is able to reproduce the data up to 1 GeV$2$, but falls short beyond this 
region.

\section{Sum rule for the integral $I_3(Q^2)$}
The only sum rule containing a prediction for the electric charge 
of the nucleon is 
\beqn
I_3(0)\;=\;\int_{\nu_0}^\infty d\nu\frac{K}{\nu}\frac{\sigma_{LT}}{Q}
\;
\begin{array}{c}
\\
\rightarrow\\
Q^2\to 0
\end{array}
\;\frac{e_N\kappa_N}{4}\;.
\eeqn

\begin{figure}
\resizebox{0.45\textwidth}{!}{%
  \includegraphics{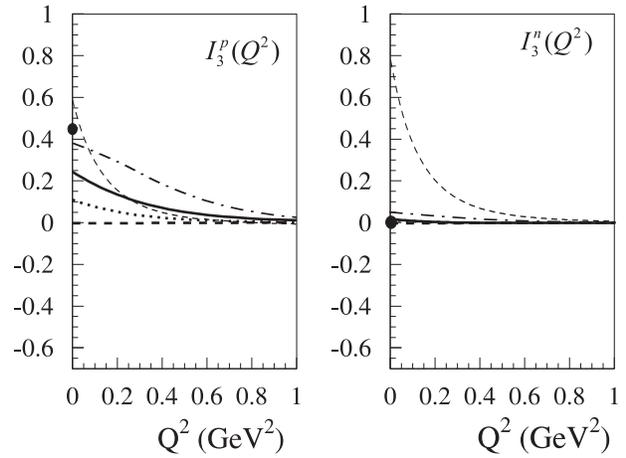}
}
\caption{Results for the $I_3$ integral for the neutron. Notation as 
in Fig. \ref{fig:baldin}. Solid cicrcle corresponds to the sum rule value.}
\label{fig:i3}
\end{figure}

In Fig. \ref{fig:i3}, we show the results for the $I_3$ integral on the 
neutron. While the MAID prediction is in complete disagreement with the 
sum rule value, one can see that the HCCQM result differs only slightly from
zero, as required by the sum rule, and HO model gives a small 
positive value. Apart from the different potential of the two CQM models, the 
presented HO calculation does not take account of the hyperfine 
mixing of the wave functions which may be responsible for the cancellation 
within this integral.

\end{document}